# Non-equilibrium Approach to Thermodynamics using Jarzynski's Equality and Diagonal Entropy

Van A. Ngo* and Stephan Haas

*Department of Physics and Astronomy, University of Southern California, Los Angeles, California 90089-0484, USA*

We combine the formalisms of diagonal entropy and Jarzynski's Equality to study the thermodynamic properties of closed quantum systems. Applying this approach to a quantum harmonic oscillator, the diagonal entropy offers a notion of temperature for closed systems away from equilibrium, and allows computing free-energy profiles. We also apply this approach to a hard-core boson lattice model, and discuss measures how to estimate temperature, entropy and measure work distribution functions. This technique offers a path to investigate the non-equilibrium quantum dynamics by means of performing work in a series of quenches.

A formalism using diagonal entropy has recently been developed to account for the thermodynamic entropy in out-of-equilibrium quantum systems [1–4]. It is based on the time average of time-dependent density operators, which makes all off-diagonal terms vanish. This time average mimics experimental measurements of physical quantities. Diagonal density operators resulting from the time average can be used to explain experimental observations that initial pure quantum states eventually evolve into mixed states. The diagonal entropy coincides with the equilibrium micro-canonical entropy in a chaotic regime, but does not converge to the entropy of a generalized Gibbs ensemble, whose system is integrable and has conserved quantities [3]. Thus, diagonal entropy can be used to evaluate temperature and other thermodynamic quantities in the chaotic regime, but is not satisfactory to do the same for integrable and closed quantum systems [3,5,6].

To measure temperature in experiments one must perturb systems sufficiently small and reach thermodynamic balance between systems and experimental instruments, which read temperature. In thermodynamics, the balance is understood as systems are imbedded in very large heat baths, thus temperature is well defined. In this picture, heat baths constantly interact with systems. In contrast, a single quench of a closed quantum system generally cannot induce the conventional thermodynamic Gibbs states [3,6,7], thus cannot model the balance between quantum systems and experimental instruments. Moreover, the use of explicit heat baths in computational studies is expensive due to their exponentially large Fock space, and thus it is elusive to examine the thermodynamic balance. In contrast, computation using diagonal entropy for integrable and closed systems is numerically achievable, and in this manuscript we examine a method to extract temperature and thermodynamic quantities from diagonal entropy.

Here we focus on thermodynamic quantities such as work and free-energy changes computed via Jarzynski's Equality (JE). Free-energy changes are used to demonstrate energetic properties of chemical reactions along pathways. Their free-energy profiles can be evaluated from out-of-equilibrium processes, whose values of work are exponentially weighted in the JE approach [8–15]. Note that it is possible to combine the JE with the formalism of diagonal entropy since both of them are applicable to out-of-equilibrium dynamics [2]. This combination offers a way to examine thermodynamic properties based on energy fluctuations in thermally isolated cyclically driven systems. Unlike other developments on the JE [8–10], which use temperature of initial thermal states to compute work distribution functions and free-energy profiles, the proposed formalism based on the JE requires that temperature is maintained along a reaction pathway of a control parameter coupled to large heat baths inducing canonical ensembles [14,15]. So far, it has not been shown how to maintain temperature along a reaction pathway in integrable and closed quantum systems, thus enabling capture of thermodynamics in chemical reactions.

This paper has three aims: (1) we present an approach mimicking the physics of measuring temperature in experiments via changing a control parameter in closed quantum systems using the formalism of diagonal entropy; (2) we demonstrate protocols of quenches to maintain temperature along a pathway of the control parameter; and (3) we show a way to combine the formalism of diagonal entropy with the JE to estimate free-energy profiles. We organize the paper as follows. Section I discusses formalism of diagonal entropy, temperature combined with Jarzynski's Equality. Section II presents two examples of applying this approach to a quantum harmonic oscillator and a lattice of hard-core bosons. Finally, in Section III we discuss the notion of temperature and conclude our findings.

**I. Theory**

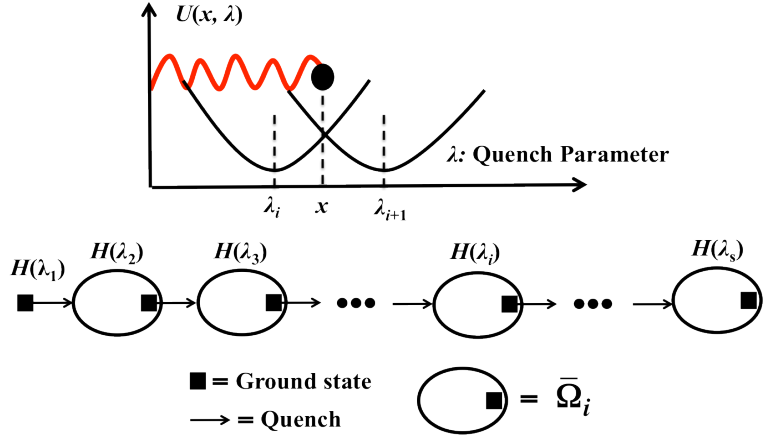

FIG. 1 (Color online). Protocol of multiple quenches to generate a set of diagonal density operators $\bar{\Omega}_i \equiv \bar{\Omega}_{\lambda_i}(\Delta\lambda)$. The parameter $\lambda$ denotes an external control, which can tune $\lambda$ at will. Each circle and square is associated with the Hamiltonian $H(\lambda_i)$. Here, the ground states are chosen as initial states for *Examples* 1 and 2, but not restricted for more general cases as long as the characteristic temperature is the same for all quenches.

A diagonal density operator can be generated in a single quench by changing a control parameter $\lambda$ along a reaction pathway with an increment $\Delta\lambda$. Given an initial state $|\psi_0(\lambda-\Delta\lambda)\rangle$ belonging to a Hamiltonian $H(\lambda-\Delta\lambda)$, at time $t \geq 0$ a perturbation is instantaneously turned on, and the system subsequently evolves with Hamiltonian $H(\lambda)$. The time-dependent density operator is then given by $\Omega(t,\lambda,\Delta\lambda) = e^{-iH(\lambda)t}|\psi_0(\lambda-\Delta\lambda)\rangle\langle\psi_0(\lambda-\Delta\lambda)|e^{iH(\lambda)t}$. A diagonal density operator is defined as

$$\bar{\Omega}_\lambda(\Delta\lambda) = \lim_{\tau\to\infty}\frac{1}{\tau}\int_0^\tau dt\, \Omega(t,\lambda,\Delta\lambda) = \sum_n |E_n\rangle\langle E_n| p_n, \quad (1)$$

where $|E_n\rangle$ is an eigenstate of $H(\lambda)$ and $p_n = |\langle E_n|\psi_0(\lambda-\Delta\lambda)\rangle|^2$. The diagonal entropy can then be defined as $S(\lambda,\Delta\lambda) = -\text{Tr}\bar{\Omega}_\lambda(\Delta\lambda)\ln\bar{\Omega}_\lambda(\Delta\lambda) = -\sum_n p_n \ln p_n$.

We assume that an infinitesimal variation of $\lambda$ does not change the thermodynamic balance between the system and experimental instruments. This assumption is consistent with the observation that heat baths constantly interact with systems to maintain temperature and other thermodynamic quantities, and any infinitesimal changes do not drive systems out of equilibrium. The inverse temperature $\beta = 1/T$ ($k_B = 1$) is computed from

$$\beta = \left(\frac{\partial S}{\partial E}\right)_\lambda \equiv \lim_{\varepsilon\to 0}\frac{S(\lambda,\Delta\lambda+\varepsilon)-S(\lambda,\Delta\lambda)}{E(\lambda,\Delta\lambda+\varepsilon)-E(\lambda,\Delta\lambda)}, \quad (2)$$

where $E(\lambda,\Delta\lambda) = \sum_n E_n p_n$ is the averaged energy. If diagonal entropy coincides with micro-canonical entropy [3], $T$ is an equilibrium temperature. If not, $T$ represents an out-of-equilibrium temperature, whose meaning will be discussed through out this paper. $T$ is the so-called a characteristic temperature associated with the diagonal density operator $\bar{\Omega}_\lambda(\Delta\lambda)$ and the diagonal entropy $S(\lambda,\Delta\lambda)$.

Now suppose that we perform a series of quenches, which yield the same characteristic temperature $T = \beta^{-1}$ [see Fig. (1)], and wish to compute free-energy profiles, $\Delta F(\lambda_1,\lambda_s)$ via Jarzynski's Equality (JE),

$$\exp[-\beta\Delta F(\lambda_1,\lambda_s)] = \int dW \rho(W)\exp(-\beta W) \equiv \langle\exp(-\beta W)\rangle, \quad (3)$$

where

$$\rho(W) = \prod_{i=1}^{s-1}\int dx_i f_i(x_i)\delta\left(W - \sum_{i=1}^{s-1}[U(x_i,\lambda_{i+1})-U(x_i,\lambda_i)]\right)$$

is the work distribution function, $\hat{U}(\hat{x},\lambda)$ is a potential operator with coupling $\lambda$ ($\lambda_i = (i-1)\Delta\lambda$) to a reaction coordinate operator $\hat{x}$, and $f_i(x_i) = \text{Tr}\bar{\Omega}_{\lambda_i}(\Delta\lambda)\delta(\hat{x}-x_i)$ is the distribution function of eigenvalue $x_i$ at the $i$-th quench. The free-energy profile $\Delta F(\lambda_1,\lambda_s)$ converges to $F(\lambda_s) - F(\lambda_1)$, where $F(\lambda_i)$ is $-\beta^{-1}\ln[\text{Tr}e^{-\beta H(\lambda_i)}]$, as the number $s$ of discretized equilibration times is taken towards infinity and the central limit theorem holds. One remarkable feature of the JE is that once small rare values of work corresponding to the optimal pathways are sampled, they are exponentially weighed, so collected data in equilibrium do not substantially change $\Delta F(\lambda_1,\lambda_s)$.

We find that such small rare values of work can be generated in the protocol illustrated in Fig. 1. This protocol is motivated by the fact that small rare values of work are available as long as the probabilities of measuring a reaction coordinate $x$ in successive quenches overlap [15]. In this protocol, the external potential $U(x,\lambda)$ is used to couple a system with an external control represented by $\lambda$. By quenching $\lambda$, we drive the system along a reaction pathway. The diagonal density operator associated with the $i$th quench and Hamiltonian $H(\lambda_i)$ is denoted by $\bar{\Omega}_i \equiv \bar{\Omega}_{\lambda_i}(\Delta\lambda)$, which generally depends on initial states. For large systems connected to heat baths, $\bar{\Omega}_i$ approaches the fully thermalized Gibbs state regardless of the initial states [16,17], and thus the requirement of the same temperature for all quenches is automatically satisfied. For small closed systems, however, one has to choose appropriate initial states for a series of quenches to have the same temperature for evaluating free-energy profiles via Eq. (3). A series of quenches characterized by a single value of $T$ and the corresponding free-energy profile can be used to quantify the thermodynamic information extracted from the quenches. We now show how to implement this technique of combining diagonal density operators with JE by simply choosing the appropriate initial states for all quenches within a sequence.

**II. Examples**

*Example* 1: Let us first consider a simple harmonic oscillator, whose Hamiltonian is given by

$$H(\lambda) = \frac{\hat{p}^2}{2m} + \frac{k\hat{x}^2}{2} + \frac{k(\hat{x}-\lambda)^2}{2} = \hbar\omega(a_\lambda^+ a_\lambda + \frac{1}{2}) + \frac{k\lambda^2}{4}, \quad (4)$$

where $\omega = (2k/m)^{1/2}$, and $a_\lambda = \sqrt{\frac{m\omega}{2\hbar}}\left(\hat{x} - \frac{j}{m\omega}\hat{p} - \frac{\lambda}{2}\right), a_\lambda^+ = \sqrt{\frac{m\omega}{2\hbar}}\left(\hat{x} + \frac{j}{m\omega}\hat{p} - \frac{\lambda}{2}\right)$ are the annihilation and creation operators. The initial state for each quench in Fig. (1) is the ground state of $H(\lambda - \Delta\lambda)$. Taking advantage of $a_\lambda = a_{\lambda-\Delta\lambda} - \sqrt{m\omega/2\hbar}\Delta\lambda/2$ and $a_{\lambda-\Delta\lambda}|n_{\lambda-\Delta\lambda} = 0\rangle = 0$, one obtains the diagonal density operator $\bar{\Omega}_\lambda = e^{-y}\sum_{n_\lambda=0}|n_\lambda\rangle\langle n_\lambda| y^{n_\lambda}/n_\lambda!$, with $y = m\omega\Delta\lambda^2/8\hbar$. In this case, the energy distribution function $p_{n_\lambda} = e^{-y}y^{n_\lambda}/n_\lambda!$ is exactly the Poisson distribution function.

One can then easily compute the averaged energy, $E = \hbar\omega(y + 1/2) + k\lambda^2/4$, and the entropy $S = y - y\ln y + e^{-y}\sum_{n_\lambda=0} y^{n_\lambda}\ln n_\lambda!/n_\lambda!$. Since $S$ is a function of $y$ only, varying $y$ (or $\Delta\lambda$) changes the entropy and energy at the same time while keeping $\lambda$ fixed. As a result, one can use the chain rule to compute the derivative in Eq. (2) as

$$\frac{1}{T} = \left(\frac{\partial S}{\partial E}\right)_\lambda = \frac{\partial S}{\partial y} / \left(\frac{\partial E}{\partial y}\right)_\lambda \quad (5)$$
$$= \frac{e^{-y}\sum_{l=1}y^l[\ln(l+1) - \ln y]/l!}{\hbar\omega}.$$

Figure 2(a) shows how $T$ increases monotonically with respect to the control parameter $y$, in comparison with $T_B$ computed from the boson distribution function $\langle n \rangle = 1/[\exp(\hbar\omega/T_B) - 1]$, where $\langle n \rangle = y$ if one takes average of the number operator $a_\lambda^+ a_\lambda$ over the density operator. For sufficiently small quenches with $y < 0.5$, $T$ is equivalent to $T_B$, and the entropy $S$ is the same as $S_B$ [see Fig. 2(b)], which is the canonical entropy of the harmonic oscillator at the same temperature $T < \hbar\omega$ maintained by heat baths. However, for large quenches, while $T$ is increasingly higher than $T_B$ at $y = \langle n \rangle$, $S$ becomes increasingly smaller than $S_B$ at the same temperature. $S$ smaller than $S_B$ is expected because large quenches produce non-equilibrium states, whose entropy is always less than the equilibrium entropy.

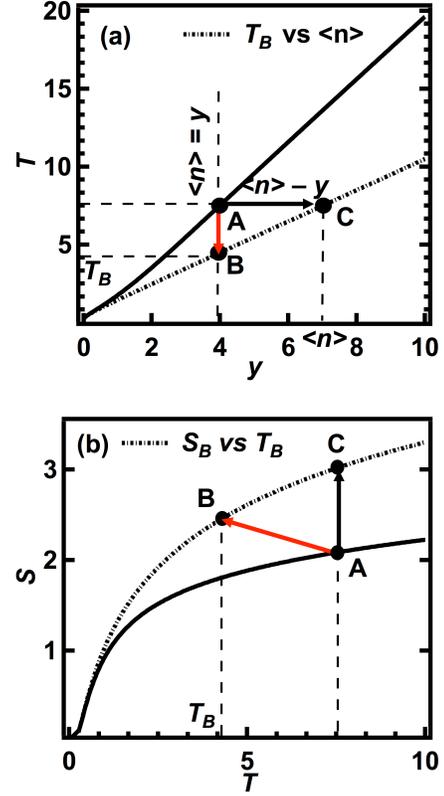

FIG. 2 (Color online). (a) Temperature $T$ (in units of $\hbar\omega/k_B$) versus $y = m\omega\Delta\lambda^2/8\hbar$ computing via Eq. (5) and (b) diagonal entropy $S$ (in units of $k_B$) versus $T$ in a quantum harmonic oscillator. $T_B$ and $S_B$ are the temperature and entropy of a non-interacting boson system computed by $\hbar\omega/\ln(1+\langle n\rangle^{-1})$ and $[(1+\langle n\rangle)\ln(1+\langle n\rangle) - \langle n\rangle\ln\langle n\rangle]$, respectively. Here, $\langle n \rangle$ is the expectation value of the number operator $a_\lambda^+ a_\lambda$. The red (A → B) and black (A → C) arrows present the transitions from an out-of-equilibrium state represented the diagonal density operator to the corresponding equilibrium states having $\langle n \rangle = y$ and $T_B = T$, respectively.

To elucidate the meaning of the out-of-equilibrium temperature $T$ being higher than $T_B$, we consider transitions from an out-of-equilibrium state A without heat baths to equilibrium states B and C, which are controlled by heat baths with $\langle n \rangle = y$ and $T_B = T$, respectively. In the first case [A → B in Fig. 2], $T \sim 2y$ higher than $T_B \sim \langle n \rangle$ at $y = \langle n \rangle$ indicates that the system reduces its temperature to $T_B$ when it is coupled to a heat bath maintained at $T_B$ to transform from the out-of-equilibrium state A to the equilibrium state B. This indicates that for large quenches the out-of-equilibrium state $\bar{\Omega}_\lambda$ is hotter than the equivalent equilibrium canonical ensemble, whose average occupation number $\langle n \rangle$ is smaller than $1/[\exp(\hbar\omega/T)-1]$, and hence $T_B < T$. In contrast, when we consider a transition at constant temperature $T_B = T$ [A → C in Fig. 2], the system's entropy is increased, and the expectation value of the number operator grows from $y$ to $\langle n \rangle$. This means that in this case the system changes its initial Poisson distribution of the energy to the canonical distribution, and increases its average energy with the entropy while the temperature remains unchanged. Indeed, the system starts absorbing energy of $\hbar\omega(\langle n \rangle - y)$ to thermalize with the heat bath for all final states in the regime between B and C. Finally, for transitions from A to final states in the regime between the origin ($y=0$, $T=0$) and B, the system emits energy $\hbar\omega(y - \langle n \rangle)$ to thermalize with the heat bath, while $S$ either increases or decreases, depending on the particular heat-bath temperature.

Now, applying the protocol described in Fig. (1), we obtain the same temperature and the same diagonal density operators for all single quenches. Given that we proceed to compute free-energy profiles. The distributions functions of $x_i$ are $f_i(x_i) = \sum_{n_{\lambda_i}=0} |\langle x_i | n_{\lambda_i}\rangle|^2 p_{n_{\lambda_i}}$, where $\langle x_i | n_{\lambda_i}\rangle$ are the wave functions of the coupled harmonic oscillator at $\lambda_i = (i-1)\Delta\lambda$. At low temperatures or small $y$, the ground states dominantly contribute to the free-energy change,

$$\Delta F(\lambda_1, \lambda_s) \approx \underbrace{\frac{k(s-1)^2 \Delta\lambda^2}{4}}_{\Delta F_{\text{Target}}} + \underbrace{\frac{k(s-1)\Delta\lambda^2}{4}(1-\frac{\hbar\omega}{2T})}_{\Delta F_{\text{Can}}} \quad (6)$$
$$+ \frac{k(s-1)\Delta\lambda^2}{4}\frac{T}{\hbar\omega}.$$

The diagonal ensemble introduces a positive shift [the last term in Eq. (6)] to $\Delta F_{\text{Can}}$, which is computed from the canonical ensembles, $f_i(x_i) = \text{Tr}\delta(\hat{x} - x_i)e^{-H(\lambda_i)/T}/\exp[-F(\lambda_i)/T]$ [15]. For a quench with $T/\hbar\omega = (-1+\sqrt{3})/2 \approx 0.37$, the last two terms cancel each other, thus $\Delta F(\lambda_1, \lambda_s)$ can be equal to $\Delta F_{\text{Target}}$. For large quenches, the last term dominates the second term, thus one expects $\Delta F(\lambda_1, \lambda_s)$ higher than $\Delta F_{\text{Target}}$.

To illustrate the contributions of excited states, we numerically evaluate the free-energy profiles for a protocol with various $\Delta\lambda$. We find that more than 50 excited states neither change temperature nor the computed free-energy profiles within an uncertainty of $10^{-6}$ for $y \in [0:10]$ because of the Poisson distribution of energy. We note that while the distribution functions $f_i(x_i)$ look Gaussian [Fig. 3(a)] for $\Delta\lambda = 0.6935$ ($y = 0.06$), they exhibit double peaks at $\Delta\lambda = 4.0$ ($y = 2$) [Fig. 3(c)]. This double-peak feature in $f_i(x_i)$ at large $y$ indicates the dominance of excited states in the Poisson distribution, which is in contrast to the always largest contributions of the ground states in canonical ensembles. It indicates that the central limit theorem breaks down in out-of-equilibrium processes where coherent or excited states become dominant [18,19]. These distributions due to large quenches are similar to those of Lohschmidt's Echo in the small-quench regime of a critical quantum XY chain [7], indicating poor equilibration. It is striking to find that the diagonal density operators can reproduce accurate free-energy profiles [see Figs. 3(b)-(d)] even in regimes, where diagonal density operators are different from canonical operators.

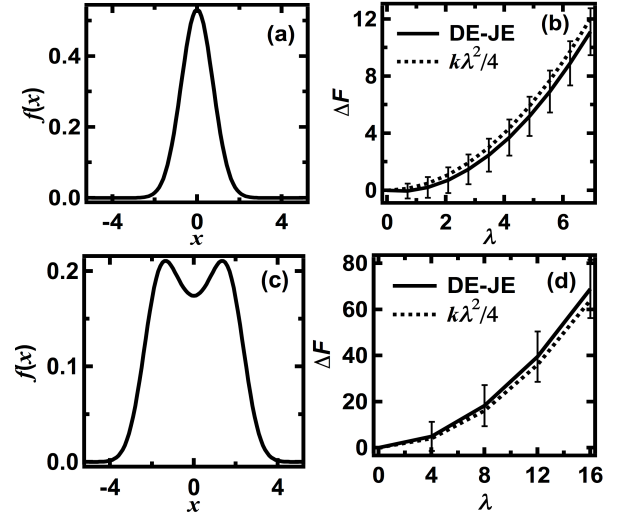

FIG. 3. (a) Distribution of reaction coordinate at $\lambda = 0$ and (b) free-energy profile at $T = 0.35$, $\Delta\lambda = 0.6935$. (c) Distribution of reaction coordinate at $\lambda = 0$ and (d) free-energy profile at $T = 3.52$, $\Delta\lambda = 4.0$. The distributions at other values of $\lambda$ are identical to (a) and (c). Free energy change $\Delta F$, $T$, and $\lambda$ are in units of $\hbar\omega/2$, $\hbar\omega/k_B$, and $\sqrt{\hbar/m\omega}$, respectively. DE-JE represents a free-energy profile computed by Diagonal Ensemble and Jarzynski's Equality. The error bars are the standard deviations of the work distribution functions.

*Example* 2: Let us now consider a one-dimensional lattice of hard-core bosons trapped in a harmonic potential [20], whose Hamiltonian after the Jordan-Wigner transformation is given by

$$H(\lambda) = -J\sum_{k=1}^{N}(f_k^+ f_{k+1} + h.c) + V\sum_{k=1}^{N} f_k^+ f_k (k-a)^2$$
$$+ V\sum_{k=1}^{N} f_k^+ f_k (k-\lambda)^2, \quad (7)$$

where $f_k$ and $f_k^+$ are fermion annihilation and creation operators at site $k$, $N$ is the number of lattice sites, $J$ is the hopping parameter, $V$ denotes the spring constant, and $a$ is a constant. We choose $V/J = 0.0225$ to obtain the superfluid phase for $N_b = 10$ hard-core bosons (equivalently 10 fermions) in a lattice of $N = 40$ sites, and $\lambda$ from $\lambda_1 = a = 13$ to $\lambda_s = 20$. We follow the method described in Ref. [20]: the initial state $|\psi_0(\lambda_i - \Delta\lambda)\rangle$ for the $i$-th quench is constructed by filling the $N_b$ lowest energy levels, whose eigenstates are obtained by diagonalizing $H(\lambda_i - \Delta\lambda)$. The excited many-body eigenstates of $H(\lambda_i)$ are constructed by assigning fermions in $H(\lambda_i)$'s ground states to the other $(N-N_b)$ energy levels. We note that for any $\Delta\lambda^2$, one must consider all many-body states that have overlaps with the initial ground states to have non-zero values of $p_n$. To estimate an effective temperature, we use Eq. (5) with entropies and energies evaluated from two quenches with $\Delta\lambda$ and $\Delta\lambda + \varepsilon$ at $\lambda = 15$ where $\varepsilon = 0.1\Delta\lambda$. The total energy $E(\lambda=15) = \langle\psi_0(\lambda-\Delta\lambda)|H(\lambda)|\psi_0(\lambda-\Delta\lambda)\rangle$ is approximately $J(0.112\Delta\lambda^2 - 0.495)$. Consistent with the results of *Example* 1, we observe that for small $\Delta\lambda^2$ the ground and a few excited many-body states overlapping with the initial states are sufficient to estimate the diagonal entropy and effective temperature. Figures 4(a) shows that $T$ is approximately proportional to $\Delta\lambda^2/8 \equiv y$ when all other parameters are made unity. This is similar to the dependence of $T_B$ on $\langle n\rangle$ for large $\langle n\rangle$ in *Example* 1. This result suggests that by increasing the number of particles, $T$ becomes closer to $T_B$, i.e., the system of ten particles appears to be a simple harmonic oscillator coupling to a heat bath. While $T$ increases slower with $y$ than $T$ in *Example* 1, Figure 4(b) shows that the diagonal entropy $S$ increases quicker with $T$ than $S$ in *Example* 1. This is consistent with fact that the entropy of ten particles should be larger than that of one particle at the same temperature.

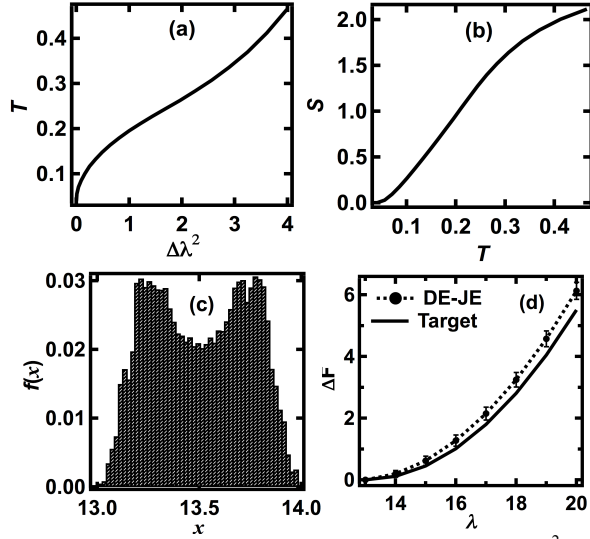

FIG. 4. (a) Temperature $T$ (in unit of $J/k_B$) versus $\Delta\lambda^2$. (b) Diagonal entropy $S$ (in units of $k_B$) versus $T$. (c) Distribution of the center-of-mass $x$ of the superfluid during the time-evolution of $|\psi(t,\lambda_i)\rangle$ following a quench with $\lambda_i = 14$. The distributions for other values of $\lambda$ are almost identical. (d) Free-energy profile in a lattice of Hard-Core bosons trapped in a harmonic potential versus $\lambda$ (*Example* 2).

The final test is to compare the free-energy profile with the target, $F(\lambda_s) - F(\lambda_1) = VN_b(\lambda_s - a)^2/2$. The work distribution functions are computed from the distributions of the center-of-mass $x_i = \sum_{k=1}^{N}\langle f_k^+ f_k\rangle_i \times k/N_b$ and the work $W = VN_b\Delta\lambda\sum_{i=1}^{s}(2\lambda_i + \Delta\lambda - 2x_i)$, where $\Delta\lambda = 1$ and $\langle...\rangle_i$ is an expectation value using time-evolving many-body state $|\psi(t,\lambda_i)\rangle = \exp(-jH(\lambda_i)t)|\psi_0(\lambda_i - \Delta\lambda)\rangle$. Here, we solve the time-evolution for the state instead of evaluating the diagonal density operators, because it is practical to approximate the distributions by

$$f_i(x_i) = \text{Tr}\delta(\hat{x} - x_i)\bar{\Omega}_{\lambda_i} \approx \lim_{\tau\to\infty}\frac{\int_0^{\tau}dt}{\tau}\delta(x_i(t) - x_i), \quad (8)$$

where $x_i(t)$ is $\langle\psi(t,\lambda_i)|\hat{x}|\psi(t,\lambda_i)\rangle$. This is a good approximation as long as

$$\lim_{\tau\to\infty}\frac{\int_0^{\tau}dt}{\tau}\int_{-\infty}^{\infty}dk\sum_n\frac{(-jk)^n}{n!}\left(\langle(\hat{x}-x_i)^n\rangle_i - [\langle\hat{x}-x_i\rangle_i]^n\right) \text{ is}$$

negligible. To obtain the convergent distributions of $x_i$ we simulate the evolution following each quench for times larger than $N^2$ [7]. Similar to the distribution in Fig. 3(c), the double-peak feature in Fig. 4(c) indicates poor thermalization for the quenches in the lattice. The characteristic temperature for these distributions is 0.1953. Using this temperature and work distribution functions in Eq. (3), we obtain the free-energy profile plotted in Fig. 4(d), which is approximately 10% higher than the target at $\lambda = 20$.

### III. Discussions and Conclusions

We have observed that the temperature computed via Eq. (2) coincides with the equilibrium temperature for sufficiently small quenches on a simple harmonic oscillator, whose diagonal entropy also converges to the canonical entropy in the non-chaotic regime. This suggests that the procedure of extracting an effective temperature $T_e$ from $\langle H\rangle = \text{Tr}[H\exp(H/k_BT_e)]/\text{Tr}[\exp(H/k_BT_e)]$, where $H$ is a many-body Hamiltonian operator of a integrable and closed system, $\langle...\rangle$ denotes an average over an initial quantum state, and Tr denotes trace, is problematic [3,6]. While $T_e$ can be incorrectly non-zero when initial states are pure, $T$ from Eq. (2) is zero if quenches represented by $(\Delta\lambda + \varepsilon)$ and $\Delta\lambda \gg \varepsilon$ change energy $\langle H\rangle$ but maintain the purity, hence diagonal entropy in systems. Moreover, *Example* 1 shows that even if $T_B$ (similar to $T_e$ if there are heat baths) is equal to $T$, the energies evaluated over a diagonal density operator and the canonical density operator can be very different. Therefore, $T_e$ should not be used to compute and compare thermodynamic quantities between diagonal and other canonical entropies.

Equation (2) suggests that information from a single quench is not sufficient to examine how thermodynamic quantities emerge in quantum mechanics, even though it offers interesting and useful information of thermalization processes in terms of time scales, system sizes, entanglement, and so on. The limit in Eq. (2) for two quenches $(\Delta\lambda + \varepsilon)$ and $\Delta\lambda$ indicates the convergence of external controls on systems. It implies the thermodynamic balance in the interactions between systems and external controls. It also implies that it is possible to model heat baths via controlling a parameter such as $\lambda$. This idea is supported by experiments of using laser-cooling techniques, in which variation of frequencies (i.e. control parameters) coupled to systems of few particles can change temperature [21].

Equation (2) offers an efficient way of estimating temperature for closed quantum systems, which might not be restricted in the framework of diagonal entropy. The most expensive computational cost for many-body systems is in evaluating the entropy $S$, and then expressing $S$ in terms of energy. Equation (2) suggests that the difference of between diagonal entropies in two quenches ($\Delta\lambda + \varepsilon$) and $\Delta\lambda$ is sufficient to compute temperature. In the two examples we find that they require a number of single-particle or many-body states as small as 50, in contrast to an exponentially large number of many-body states in Fock space to accurately compute a single value of $T$.

*Example* 1 suggests one interesting effect that even if $T$ computed for the diagonal density operator is equal to the temperature of heat baths, the system of a simple harmonic oscillator can still absorb a net energy to thermalize with heat baths by redistributing occupancy probabilities over energy levels, thus increasing entropy. This effect suggests a way to experimentally verify the temperature by measuring if the fluctuations of the heat-bath temperature before and after coupling with an out-of-equilibrium system are the same, while the system absorbs a significant energy ($\Delta E > 0$). Note that according to the approach, the opposite effect with a net energy emission ($\Delta E < 0$) cannot occur while there is no significant change in the fluctuations of temperature during the system and heat-bath coupling, because the entropy increases ($\Delta S > 0$).

In *Example* 2, we observe that the integrable and closed system of ten particles has temperature computed via Eq. (2) similar to the equilibrium temperature of a harmonic oscillator coupled to heat baths. This further confirms the procedure of computing temperature via Eq. (2), which yields the equilibrium temperature as a number of particles increases [16]. This procedure provides a generalized notion of temperature more general than the traditional Gibbs construction, which requires a large number of particles and chaotic collisions for thermodynamic equilibrium.

This approach also offers a way to examine thermodynamics in chemical reactions, which are prototypical out-of-equilibrium processes. One is able to compute temperature via Eq. (2) to model and characterize the thermodynamic balance between systems and surroundings, whose direct computation is expensive. Estimates of free-energy profiles from out-of-equilibrium processes of chemical reactions via the diagonal entropy and the JE can be used to generate energetic diagrams describing how chemical reactions may occur. We are currently attempting to apply this approach to examine chemical reactions in testable quantum systems such as hydrochloric acid and water molecules.

In conclusion, we have presented an approach combining Jarzynski's Equality with diagonal entropies to out-of-equilibrium closed quantum systems. We tested this approach on a harmonic oscillator and on a lattice of hard-core bosons verifying that even though diagonal ensembles can be different from canonical ensembles, it is possible to extract meaningful thermodynamic quantities such as temperature, work and free-energy profiles.

**Acknowledgements:** We would like to acknowledge useful conversations with G. E. Crooks, T. Albash, and L. C. Venuti, and financial support by the Department of Energy grant DE-FG02-05ER46240.